\documentclass[aps,twocolumn,amsfonts,amssymb,showpacs]{revtex4}

\usepackage{graphicx,bm}
\usepackage{amsmath}

\newcommand{\be}{\begin{equation}}
\newcommand{\bea}{\begin{eqnarray}}
\newcommand{\ba}{\begin{array}}
\newcommand{\ee}{\end{equation}}
\newcommand{\eea}{\end{eqnarray}}
\newcommand{\ea}{\end{array}}
\newcommand{\nn}{\nonumber}

\begin{document}

\title{Schwinger model on a half-line}

\author{Yeong-Chuan Kao and Yu-Wen Lee }
\affiliation{ Department of Physics, National Taiwan University,
Taipei, Taiwan}
\date{Received 7 August 2001}
\begin{abstract}
We study the Schwinger model on a half-line in this paper. In
particular, we investigate the behavior of the chiral condensate
near the edge of the line. The effect of the chosen boundary
condition is emphasized. The extention to the finite temperature
case is straightforward in our approach.

\end{abstract}
 \maketitle

\section{Introduction}

The (1+1)-dimensional massless spinor electrodynamics, the Schwinger
model~\cite{Sch}, has been a most popular playground for theorists because it exhibits many
subtle properties which are believed to exist, yet difficult to verify exactly, in four-
dimensional QCD~\cite{Low,manton,book}. One such property is the existence of the chiral
condensate, an essential component in modern particle physics. In the Schwinger model, a
rather complete understanding of the chiral condensate under various circumstances, e.g., at
finite temperature and finite chemical potential~\cite{nelson,kao1,wipf,smilga,Hosotani,Kao},
has been achieved. It is the unrealistic low dimensionality of the model that allows such
detailed understanding. But sometimes realistic low-dimensional models can arise from
(3+1)-dimensional ones if only radial dependence ($s$-wave approximation) is kept. For example,
the famous Callan-Rubakov effect, i.e., the monopole catalysis of proton decay
~\cite{Callan,Rubakov,mono}, made use of a Schwinger-like model defined on a half-line in the
first approximation. A fermion number breaking condensate analogous to the chiral condensate
can be shown to exist around a magnetic monopole.

Motivated partly by the
Callan-Rubakov effect, we study the original Schwinger model on a half-line. We pay
special attention to the dependence of the chiral condensate on the distance from the edge.
Exact results can be obtained by using existing techniques. The standard value of the chiral
condensate is recovered when we are in a region far away from the edge. The value of the
condensate near the edge depends on the chosen boundary condition. In this paper, we adopt the
boundary condition employed in the boundary conformal field theory approach to the
Callan-Rubakov effect. It is easy to extend the result to the finite temperature case with
our approach. Previous attempts to study the finite temperature behavior of the
Callan-Rubakov condensate relied on the cluster decomposition property which is hard to implement
on a half line~\cite{kao}.
A recent work
by D\"{u}rr \cite{durr}, building on and correcting ealier results by D\"{u}rr and
Wipf \cite{wipf2,durr} has studied the behavior of the chiral condensate in
the Schwinger model
on a finite segment with boundary conditions similar to ours. D\"{u}rr's result is
the same as ours in regions where comparisons can be made,
yet his approach and emphasis is rather different from those adopted here.

\section{Bosonization of QED$_2$ on a Half-Line}

We shall begin by first studying the Schwinger
model on a finite segment. The formulation we have followed is rather standard~\cite{book}.
 The Lagrangian density of QED$_2$ on a line segment of length
$L$ ($x\in[0,L]$) is defined by
\bea {\cal L}=-\frac{1}{4}F_{\mu\nu}^2+\bar{\psi}\gamma_\mu(i
\partial_u-eA_\mu)\psi. \eea The boundary condition on $\psi$ is chosen to be
$\psi_R(t,L)+\psi_L(t,L)=0=\psi_R(t,0)+\psi_L(t,0)$
which breaks chiral symmetry at $x=0$ and $L$. \cite{mono}
A similar boundary
condition appears in the Callan-Rubakov effect and allows an exact treatment of the theory.
Discussions on a generalization of this boundary condition will be given in Sec. IV.

The half-line case will be obtained if we
take the infinite $L$ limit at the end of the analysis.
In two dimensions, it is possible to choose the Coulomb gauge so that $A_{x}$ is independent
of the spatial coordinate
$x$, i.e. $\partial_x A_x=0$. In the Coulomb gauge, the time component of the
gauge field $A_0$ is obtained as
$A_0(t,x)=-e\int_0^L dy D(x,y)j^0 (t,y)$
by solving
the Gauss's law
\bea
E^\prime=ej^0, \qquad
\dot{E}=-ej^1.
\eea
$D(x,y)$ is the
Green's function which satisfies $\partial_x^2
D(x,y)=\delta(x-y)$.
To further our analysis, we will restrict ourselves in the
charge zero sector, i.e. $Q\equiv\int_0^L dx j_0(x)=0$.
This immediately implies the boundary condition for the electric field: $E(0)=E(L)$.
The form of the Green function $D(x,y)$ obeying the restriction is found to be
\bea
D(x,y;\lambda)=\frac{1}{2}|x-y|+a(x+y)+c
\eea
if we require $A_0(x)$ to obey the following boundary condition :
$ (d/dx)A_0 (0,t)=0= (d/dx)A_0 (L,t)$. \cite{BC}
Here $a$ and $c$ are arbitrary constant.

We define $(eL/2\pi) A_x=W(t)$, the Wilson line operator,
which will be needed in later formulations.
It is perhaps worthwhile to emphasize that as our theory is not defined on a torus (our
boundary conditions are "open"), the winding number
for the gauge field is not defined and the Wilson line operator is not an angular variable,
 due to the absence of topologically nontrivial gauge
transformations~\cite{wipf2}.

Following standard procedures, the
Hamiltonian reads
\bea
H&=&\frac{F^2}{2L}+\int_0^L
[\psi_R^\dagger i\partial_x \psi_R(x)- \psi_L^\dagger i\partial_x
\psi_L(x)]\nn\\
&+&\frac{2\pi }{L}W(t)\int_0^L dx[ \psi_R^\dagger
\psi_R(x)-\psi_L^\dagger\psi_L(x)]\nn\\
&-&\frac{e^2}{2}\int_0^L
dx\int_0^L dy j^0(t,x)D(x,y)j^0 (t,y), \label{hamil}
\eea
where
$j^0(t,x)\equiv \psi_R^\dagger \psi_R(x)+\psi_L^\dagger\psi_L(x)$.
The fermion field is written in terms of
right- and left-moving components explicitly for later
convenience.

The Hamiltonian~(\ref{hamil}) is essentially a model of fermions
interacting via long-range Coulomb forces together with
the quantum mechanical degrees of freedom  of the spatial
component of
the gauge field. We will empoly the methods
developed for boundary critical
phenomena\cite{affleck,polchinski}.

The first step of bosonizing~\cite{Hosotani,Shankar} the fermions with the chosen boundary
conditions at the origin is mirror copying: our boundary condition
$\psi_R(t,0)=-\psi_L(t,0)$ allows us to analytically continue the
left-moving component of the fermion field to the domain $x<0$ by
defining $\psi_R(t,x)\equiv-\psi_L(t,-x)$ for $x\in[-L,0]$. Our
theory can then be formulated as a theory of a chiral (right-moving)
fermion defined on the full segment ($x\in[-L,L]$) which leads to
the full line in the infinite $L$ limit. Our Hamiltonian then
reads as \bea H&=&\frac{F^2}{2L}+\int_{-L}^L dx \psi_R^\dagger
i\partial_x \psi_R(x) \nn\\&+&\frac{2\pi }{L}W(t)\int_{-L}^L dx~
\mbox{sgn} (x) \rho_R(x)\nn\\&-&\frac{e^2}{2}\int_0^L dx\int_0^L dy
[\rho_R(x)+\rho_R(-x)]D(x,y)\nn\\&\times&[\rho_R(y)+\rho_R(-y)],
\label{hamilR}
\eea
where
$\rho_R(x)\equiv\psi_R^\dagger(x)\psi_R(x)$. We now bosonize the
Hamiltonian by introducing a single chiral boson field
$\hat\phi(x)$ with $\psi_R(x)=(1/\sqrt
{2L}):e^{i\hat\phi(x)}:=(1/\sqrt {2\pi a})e^{i\hat\phi(x)}$,
where $a$ is an ultraviolet cutoff~\cite{Hosotani,Shankar}. Notice that our
boundary condition on $x=L$ , $\psi_R(t,L)=-\psi_L(t,L)$, now
implies that we have the periodic boundary conditions for the
chiral fermion, i.e., $\psi_R(L)=\psi_R(-L)$. The chiral boson
therefore has the following mode expansion : \bea
\hat\phi(t,x)&=&q_R+\frac{2\pi p_R (t-x)}{2L}+\phi(t,x),\nn\\
\phi(t,x)&=&\sum_{q>0 }
\sqrt{\frac{\pi}{qL}}(e^{iqx-aq/2}b_q+ \mbox{H.c.}),
\eea
 $q=(2\pi n/2L)$ and the boson creation
operators satisfy the usual commutation relations $[b_q,
b_{q^\prime}^\dagger]=\delta_{q,q^\prime}$, $[q_R,p_R]=i$.
The periodic boundary condition requires $p_R$
to take integer values only. The meaning of various factors in the above
formulas can be seen from the following bosonization identity:
\bea
\psi_R^\dagger(x)\psi_R(x)&=&\frac{1}{2\pi}\partial_x
\hat{\phi}(x)=\frac{p_R}{2L}+\frac{\partial_x \phi(x)}{2\pi},\nn\\
\psi_L^\dagger(x)\psi_L(x)&=&\frac{-1}{2\pi}\partial_x
\hat{\phi}(-x)=\frac{p_R}{2L}-\frac{\partial_x \phi(-x)}{2\pi},
\eea
so that $p_R=\int_0^L dx(\psi_R^\dagger
\psi+\psi_L^\dagger\psi_L)$ is the charge operator while the
chiral charge operator is $\int_0^L dx [\psi_R^\dagger
\psi_R-\psi_L^\dagger\psi_L]=(1/\pi)[\phi(L)-\phi(0)]\equiv
q_5$. Notice that the chiral charge operator here appears as a surface term rather than the
zero mode of Dirac fermions. This is because the theory is defined
on a segment rather on a torus.

Some straightforward manipulations bring us the bosonized kinetic energy term
\bea
H_F=\frac{1}{4\pi}\int_{-L}^L dx \left(\partial_x \phi +\frac{2\pi
p_R}{2L} +\frac{2\pi}{L}W \mbox{sgn}(x)\right)^2.
\eea
While for the interaction term, we have
\bea
\label{Hint}
H_{int}&=&\frac{e^2}{8\pi^2}\int_0^L dx
[\phi(x)-\phi(-x)]^2\nn\\&=&\frac{e^2}{8\pi^2}\int_{-L}^L dx
[\phi(x)^2-\phi(x)\phi(-x)],
\eea
which is manifestly gauge invariant.

Since we have restricted ourselves to the charge zero sector,
$p_R$ is set to zero.
Setting everything together, we finally arrive at the following bosonized Hamiltonian :
\begin{align}
H&=\frac{F^2}{2L}+
\frac{2\pi Q_5^2}{L} \nn\\
&+\frac{1}{4\pi}\int_{-L}^L dx\left[[\partial_x \phi(x)]^2
+\frac{m^2}{2}[\phi(x)^2-\phi(x)\phi(-x)]\right]\nn\\
&=\frac{F^2}{2L}+
\frac{2\pi Q_5^2}{L}\nn\\
& +\sum_{q>0}\frac{m^2/2
+q^2}{q}\left[b_q^\dagger b_q-\frac{m^2/2}{2(m^2/2+q^2)}(b_q b_q
+b_q^\dagger b_q^\dagger)\right], \label{bosonHam}
\end{align}
where $m^2=e^2/\pi$ and $Q_5\equiv 2W-q_5$. 
As the Hamiltonian contains only quadratic terms, we can solve for
the ground state wave functional exactly.
Terms of order of $O(1/L)$ in the Hamiltonian will be
neglected from now on as they will not influence the final result.
We therefore concentrate on terms propotional to $b_q b_q$,
$b_q^\dagger b_q^\dagger$, or $b_q^\dagger b_q$. This part of the
Hamiltonian can be diagonalized easily via a Bogoliubov
transformation
\bea
b_q&\equiv&\cosh \varphi_q~ \hat{b}_q-\sinh\varphi_q~ \hat{b}_q^\dagger\nn\\
&&\mbox{with}\qquad \tanh 2\varphi_q=\frac{-m^2/2}{m^2/2+2q^2}.
\label{bogo}
\eea

The diagonalized Hamiltonian now reads
\bea
H=\sum_{q>0}
\sqrt{m^2+q^2} ~\hat{b}_q^\dagger\hat{b}_q.
\eea

This shows clearly the well known fact that the spectrum contains a relativistic boson with
mass $m$. The spectrum contains only a positive momentum branch reflecting the fact that $\phi$
is a chiral boson. With this simple bosonic Hamiltonian and the bosonic
representations of fermionic operators, we can easily calculate the fermion
correlation functions.

\section{Chiral condensate}

The chiral condensate is given by
$\bar\psi(x)\psi(x)=\psi_R^\dagger(x)\psi_L(x)+ \mbox{H.c.}$ When written
in terms of the chiral bosonic field it becomes : 
\bea
&&\psi_R^\dagger(x)\psi_L(x)=-\psi_R^\dagger(x)\psi_R(-x)\nn\\&=&\frac{-1}{2\pi
a}e^{-i\hat{\phi}(x)}e^{i\hat{\phi}(-x)}=\frac{-1}{2\pi
a}:e^{-i\phi(x)+i\phi (-x)}:\nn\\
&&\times~ e^{\langle
0\mid\phi(x)\phi(-x)-\frac{1}{2}[\phi(x)\phi(x)+\phi(-x)\phi(-x)]\mid
0\rangle}.\label{normal}
 \eea
 To proceed , we need the following relations from
Eq.~(\ref{bogo}): \bea \langle 0\mid b_q b_{q^\prime}\mid
0\rangle&=&-\cosh\varphi_q\sinh\varphi_{q^\prime} \langle 0\mid
{\hat b_q}{\hat b_{q^\prime}}^\dagger \mid
0\rangle\nn\\&-&\sinh\varphi_q
\cosh\varphi_{q^\prime}\langle 0\mid {\hat b_q}^\dagger {\hat b_{q^\prime}}\mid 0\rangle ~~,\nn\\
\langle 0\mid b_q b_{q^\prime}^\dagger\mid 0\rangle&=&\cosh\varphi_q\cosh\varphi_{q^\prime}
\langle 0\mid {\hat b_q}{\hat b_{q^\prime}}^\dagger \mid 0\rangle\nn\\&+&\sinh\varphi_q
\sinh\varphi_{q^\prime}\langle 0\mid {\hat b_q}^\dagger {\hat b_{q^\prime}}\mid 0\rangle ~~, \nn\\
\langle 0\mid b_q^\dagger b_{q^\prime}\mid 0\rangle&=&\sinh\varphi_q\sinh\varphi_{q^\prime}
\langle 0\mid {\hat b_q}{\hat b_{q^\prime}}^\dagger \mid 0\rangle\nn\\&+&\cosh\varphi_q
\cosh\varphi_{q^\prime}\langle 0\mid {\hat b_q}^\dagger {\hat b_{q^\prime}}\mid 0\rangle.
\label{relation}
\eea
With the above relations, we obtain
\bea
&&D_1 (x)=\langle 0\mid\phi(x)\phi(-x)\mid 0\rangle\nn\\&=&\sum_{q>0}\frac{\pi}{qL}
\left[-\sinh 2\varphi_q +\cosh 2\varphi_q \cdot \cos 2qx +i\sin 2qx\right] ~~,\nn\\
&&D_2 (x)=\langle 0\mid\phi(x)\phi(x)\mid 0\rangle\nn\\&=&\sum_{q>0}\frac{\pi}{qL}
\left[-\sinh 2\varphi_q \cdot \cos 2qx+\cosh 2\varphi_q \right] ~~, \nn\\
&&D_1 (x)-D_2 (x)
\nn\\&=&\sum_{q>0}\frac{\pi}{qL}\frac{q}{\sqrt{m^2+q^2}}\left(\cos
2qx -1\right)+ \sum_{q>0}\frac{\pi}{qL}\cdot i \sin 2qx.
\nn\\&&
\eea
Using the following mathematical formulas
\bea
\int_0^\infty dq \frac{\cos 2qx}{\sqrt{m^2+q^2}}&=&K_0 (2mx),\nn\\
-\hspace{-0.5cm}\sum_{q=n\pi/L,n>0}\frac{e^{-aq}}{\sqrt{m^2+q^2}}&=&{\cal \gamma}+\ln
\frac{ma}{2}+O\left(\frac{1}{L}\right),\nn\\
\sum_{q>0}\frac{\pi}{qL}e^{iqz-\alpha
z}&=&\ln\frac{L}{\pi\alpha}+\ln\frac{\alpha}{\mid
z\mid}+\frac{i\pi}{2}sgn(z),\nn\\&& \label{appro}
\eea
where $K_0$ is the modified Bessel function and $\cal\gamma$ is the Euler
constant, we find
\bea
\frac{-1}{2\pi a} e^{D_1(x)-D_2
(x)}=\frac{m}{4\pi}e^{\cal\gamma}\cdot e^{K_0(2mx)}.
\eea

Putting everything together, we finally get
\bea
\langle\bar\psi(x)\psi(x)\rangle=
\frac{m}{2\pi}e^{\cal\gamma}\cdot
e^{K_0(2mx)}. \label{T=0}
\eea
The above formula is
obtained for $L\rightarrow \infty$. The right edge of the segment at $L$ is pushed
away to infinity, so we have effectively the behavior of the condensate on a half-line.
When
$x$ is near the origin, the condensate becomes singular
because of the chiral symmetry breaking boundary condition and the exact correlation
of $\psi_L$ and $\psi_R$ at the origin.
 In regions far away from the origin, we expect that the condensate is not affected by
the boundary condition. Indeed, when
$x\gg m$,
$K_0(mx)\rightarrow 0$ and $\bar\psi(x)\psi(x)\rightarrow
(m/2\pi)e^{\cal\gamma}$ which is just the usual chiral condensate
 on an infinite line as it should be~\cite{nelson,manton,kao1,wipf,smilga,Hosotani}.
 When $x$ is very near the edge, the condensate diverges as $(1/x)$. This
behavior of the condensate near the boundary is
 also found by D\"{u}rr \cite{durr}.

The above result can be readily generalized to the finite
temperature case. One only need to substitute $\langle {\hat
b_q}^\dagger {\hat b_q}\rangle=1-\langle {\hat b_q} {\hat
b_q}^\dagger \rangle =n(T)\equiv 1/({e^{\omega(q)}-1})$ into
eq.~(\ref{relation}). Here $\omega(q)=\sqrt{m^2+q^2}$ is the
energy dispersion for the boson excitations. Upon doing that, we
have
\begin{align}
D_1(q)-D_2(q)&=\sum_{q>0}\left\{[1+2n(T)](\cosh
2\varphi_q+\sinh 2\varphi_q)\right.\nn\\
&\times \left.(\cos 2qx-1)+i\sin 2qx\right\},
\end{align}
and the chiral condensate at finite
temperature becomes
\begin{align}
&\langle\bar\psi(x)\psi(x)\rangle_{T>0}\nn\\
=&\langle\bar\psi(x)\psi(x)\rangle_{T=0}\times
\exp{\sum_{q>o}\frac{\pi}{L}\frac{2}{e^{\sqrt{m^2+q^2}/T}-1}\frac{\cos 2qx}
{\sqrt{m^2+q^2}}}\nn\\ \times
&\exp{\sum_{q>0}\frac{\pi}{L}\frac{-2}
{e^{\sqrt{m^2+q^2}/T}-1}\frac{1}{\sqrt{m^2+q^2}}}.
\label{finiteT}
\end{align}

For $mx\gg 1$, the second factor approaches $1$ and the above result
again coincide with the one obtained in the whole line case.
Equations (\ref{T=0}) and (\ref{finiteT}) are the main results of
this paper. Inclusion of a finite chemical potential is also
straightforward following a previous treatment for the standard case \cite{Kao}.

\section{Discussions}

It is important to point out that because of the symmetry breaking
boundary condition, a unique ground state has been selected among
the usual degenerate vacua far away from the boundary.
 The analogous
situation in the usual spin system with a boundary is familiar: If
we impose the condition that spins on the boundary must all point
to the same direction, say up, the spin system will settle down to
a state with spins all pointing up when temperature lower than the
critical temperature is approaching zero; one unique ground state
is selected. If we rotate the boundary spin, the orientation of
the spins deep inside will also change accordingly. Likewise here,
if we change the boundary condition to
 $\psi_R(t,0)=-e^{i\theta}\psi_L(t,0)$, the chiral condensate
Eq.~(\ref{T=0}) and Eq.~(\ref{finiteT}), will contain an extra
factor $\cos(\theta)$ where $\theta$ is the famous vacuum angle. To
understand this, one only needs to look at eq.~(\ref{normal}) and
see that an extra factor $e^{i\theta}$ must be put on the right-hand side.
The $\cos(\theta)$ factor arises from combining
$e^{i\theta}$ with $e^{-i\theta}$ that will thereby appear in the
Hermitian conjugate of Eq.~(\ref{normal}).

Finally, we note that for the free massless fermion case, i.e., charge $e=0$,
the chiral condensate can be easily calculated
from Eq. (\ref{normal}) to be proportional to $1/{2\pi x}$ when $L\to\infty$.
As a check, we can also obtain this result by taking the limit $m(=e/{\sqrt\pi})\to 0$
 in Eq.~(\ref{T=0}).

Summing up, we have demonstrated that the chiral condensate can be
obtained exactly in the Schwinger model on a half-line if we
choose an appropriate (yet not unphysical) boundary condition.
More general boundary conditions may not allow exact treatments.

 \acknowledgments

 This work is supported by the National Science Council of Taiwan under
 grant NSC 89-2112-M-002-056.


\begin{thebibliography}{99}
\bibitem{Sch} J. Schwinger, Phys. Rev. {\bf 128}, 2425 (1962).
\bibitem{Low} J.H. Loewenstein and J.A. Swieca, Ann. Phys. (N.Y.) {\bf 68}, 172 (1961) .
\bibitem{manton} N. S. Manton, Ann. Phys. (N.Y.){\bf 159}, 239 (1985)
\bibitem{book} E. Abdalla, M. C. B. Abdalla, and K. D. Rothe, {\em Non-perturbative
Methods in Two-Dimensional Quantum Field Theories} (World Scientific, Singpore, 1991).
\bibitem{nelson} N. K. Nelson and B. Schroer, Nucl. Phys. {\bf B210}, 62 (1977).
\bibitem{kao1} Y.-C. Kao, Mod. Phys. Lett. {\bf A} {\bf 7}, 1411 (1992).
\bibitem{wipf} I. Sachs and A. Wipf, Helv. Phys. Acta {\bf 65}, 652 (1992).
\bibitem{smilga} A. V. Smilga, Phys. Lett. {\bf B 278}, 371 (1992).
\bibitem{Hosotani}J. E. Hetrick and Y. Hosotani, Phys. Rev. D {\bf 50}, 2621 (1988);
J. E. Hetrick, Y. Hosotani, and S. Iso, {\em ibid.} {\bf 53}, 7255 (1996).
\bibitem{Kao}Y.-C. Kao and Y. W. Lee, Phys. Rev. D {\bf 50}, 1165 (1994).
The unusual oscillatory behavior in the chiral condensate when
chemical potential is not zero is due to the momentum anomaly
explained by Manton in Ref.~\cite{manton}, 229.
\bibitem{Rubakov}V. A. Rubakov, Nucl Phys {\bf B203}, 311 (1982).
\bibitem{Callan} C. Callan, Phys. Rev. D {\bf 25}, 2141 (1982).
\bibitem{mono}I. Affleck and J. Sagi, Nucl. Phys. {\bf B417},
413 (1994); J. Polchinski, {\em ibid.} {\bf B242}, 345 (1984).
\bibitem{kao} Y.-C. Kao, Phys. Lett. {\bf B 143}, 147 (1984).
\bibitem{durr} Stephan D\"{u}rr, Ann. Phys. (N.Y.) {\bf 273}, 1 (1999).
\bibitem{wipf2} Stephan D\"{u}rr and Andreas Wipf, Annals Phys. {\bf 255}, 333 (1997)
\bibitem{BC} This boundary condition on $A_0$ will allow us to avoid nonessential infrared
           problems.
\bibitem{affleck}S. Eggert and I. Affleck, Phys. Rev. B {\bf 46}, 10866 (1992);
M. Fabrizio and A. Gogolin, {\em ibid.} B {\bf 51}, 17825 (1995).
\bibitem{polchinski}J. Polchinski and L. Thorlacius, Phys. Rev. D {\bf 50}, R622 (1994).
\bibitem{Shankar} For a concise review of bosonization technique, see also,
R. Shankar, Lectures given at the BCSPIN School, Katmandu, 1991,
in {\em Condensed Matter and Particle Physics}, edited by
Y. Lu, J. Pati, and Q. Shafi (World Scientific Singapore 1993).




\end{thebibliography}
\end{document}